\documentclass[prl,twocolumn,floatfix,preprintnumbers,letterpaper]{revtex4}
\usepackage{graphics}
\usepackage{epsfig}
\usepackage{amsmath}

\def\al{\alpha}
\def\be{\beta}

\begin{document}

\title{A New Approach to Cosmological Bulk Viscosity}
\author{Marcelo M. Disconzi}
\affiliation{Department of Mathematics, Vanderbilt University,
Nashville, TN  ~~37235}
\author{Thomas W. Kephart}
\affiliation{Department of Physics and Astronomy, Vanderbilt University,
Nashville, TN  ~~37235}
\author{Robert J. Scherrer}
\affiliation{Department of Physics and Astronomy, Vanderbilt University,
Nashville, TN  ~~37235}
\date{\today}

\begin{abstract}
We examine the cosmological consequences of an alternative to the standard expression for bulk viscosity, one
which was proposed to avoid the propagation of superluminal signals without
the necessity of extending the space of variables of the theory.
The Friedmann equation is derived for this case, along with an
expression for the effective pressure. We find solutions  for the evolution
of the density of a viscous component, which differs markedly from the case of conventional Eckart theory;
our model evolves toward late-time phantom-like behavior with a future singularity.
Entropy production is addressed,
and some similarities and differences to approaches
based on the Mueller-Israel-Stewart theory are discussed.
\end{abstract}

\maketitle

\section{I. Introduction}

The standard expression for relativistic viscosity was derived by
Eckart in 1940 \cite{Eckart}, and cosmological implications of the Eckart viscosity were examined by Treciokas and Ellis
\cite{TE} and Weinberg \cite{Weinbergpaper,Weinbergbook}, and subsequently by several others \cite{Murphy,HZS,HS}.
Later the possibility of bulk viscosity was explored in the context of inflation
\cite{Barrow1,Waga,Pad,Barrow2,Barrow,Barrowbook,Gron} and as
a source for the accelerated expansion of the universe
\cite{Zimdahl,Balakin,Colistete,Avelino,Hipolito1,Hipolito2,Gagnon,Velten}, including
the possibility that a single fluid with viscosity could account for both
 dark matter and  accelerated expansion,
although the latter idea faces severe
difficulties \cite{LiBarrow,VeltenSchwarz}.  For a recent review of cosmological
bulk viscosity, see Ref. \cite{BG}.

It is well-known that the Eckart expression for viscosity has the flaw that it can yield superluminal signal
propagation.  Various proposals have been put forward to remedy this problem \cite{Mueller,Israel,IsraelStewart,Geroch,Carter},
with the most widely-studied being the Mueller-Israel-Stewart (MIS) 
theory of Refs. \cite{Mueller,Israel,IsraelStewart}.  Cosmological implications
of the latter are examined  in 
Refs. \cite{PBJ,Maartens,Piatella}, and further discussion can be found in Refs. 
\cite{BB, NO, Dos_et_al,DosTsa,Bastero-Gil:2014jsa,
Pahwa_FLRW, KSK, BreGor}.

Here we discuss  cosmological aspects of a more recent proposal to evade the causality issue, namely, the model introduced by
Disconzi \cite{Disconzi}, based on earlier work by Lichnerowicz 
\cite{Lichnerowicz, LichnerowiczBook},
and generalized in Ref. \cite{DisCzu}.  
In the Eckart theory, the bulk viscosity
is derived from the divergence of the four-velocity of the fluid, i.e., the stress-energy tensor is
given by (we take $c = 8 \pi G = 1$ throughout):
\begin{equation}
T^E_{\alpha\beta} = (p+\rho) u_\al u_\be + p g_{\al\be}
- \zeta (g_{\al \be }+ u_\al u_\be ) \nabla_\mu u^\mu,
\label{T_N}
\end{equation}
where $p$ is the pressure, $\rho$ is the density, $u$ is the four-velocity,
and $g$ is the metric (with
convention $-+++$). The viscosity coefficient, $\zeta$, 
is not necessarily constant, and is frequently taken
to vary as an unknown power of the fluid density.
Since Eq. (\ref{T_N}) can lead to superluminal signals, Disconzi \cite{Disconzi} proposed
instead that
\begin{gather}
T_{\alpha\beta} = (p+\rho) u_\al u_\be + p g_{\al\be}
- \zeta (g_{\al \be }+ u_\al u_\be ) \nabla_\mu C^\mu,
\label{T_vis}
\end{gather}
where $C$ is the dynamic velocity, defined by 
$C_\al = F u_\al$ \cite{LichnerowiczBook}. $F$ is called
 the index of the fluid (see below) and depends on the nature of the fluid
(and thus on the equation of state).
We can think of $F$ as providing a suitable 
relativistic correction (see Sec. V) to the formulation of the velocity in 
the viscous case, 
as the very definition of the velocity four-vector
is somewhat ambiguous when viscosity is present \cite{RZ}. On the other
hand, this ambiguity is absent  when $\zeta = 0$; hence, it is plausible
to introduce a modification only in the viscous part of Eq.
(\ref{T_N}), leading to Eq. (\ref{T_vis}).

The model for viscosity from Refs. \cite{Disconzi,DisCzu} leads
to a well-posed theory without superluminal signals under
many interesting conditions.  While it has not been rigorously proven
to be causual under all possible circumstances, there are no known
systems in which it is non-causal, so it is plausible to conjecture
that it is causal under all conditions.  In comparison,
the Eckart model for viscosity can be shown to be non-causal under
some conditions.  The MIS model has the same status as the model under
consideration here, namely, it has not been rigorously proven
to be causal, but there are no known systems for which it is non-causal.
Hence, in terms of causality, the model of Refs. \cite{Disconzi,DisCzu}
is at least an improvement on the Eckart theory, and it is no worse
than the MIS model, while being much simpler than the latter.

The model under consideration here \cite{Disconzi,DisCzu} uses 
\begin{gather}
F = \frac{p+\rho}{\mu},
\label{def_F_general}
\end{gather}
where $\mu$ is the rest mass density, 
although in principle other choices of $F$ could be explored.
Recall that $\mu$ is conserved
along the flow lines, i.e., 
\begin{gather}
\nabla_\al (\mu u^\al) = 0.
\label{mu_conservation}
\end{gather}

Now we briefly turn our attention to some technical points.
While it might appear from Eq. (\ref{T_vis}) that
$F$ can be rescaled by an arbitrary constant, with the constant
absorbed into the definition of $\zeta$, leaving $T_{\alpha\beta}$
unchanged, this is, in fact, not the case.  The quantity $F$ is not
an arbitrary parameter, but is the specific enthalpy given by Eq.
(\ref{def_F_general}). Further, $F$ is not, in general, constant
except for certain special cases.

Although Eq. (\ref{T_vis}) reduces to Eq. (\ref{T_N}) upon setting
$F=1$, which is useful for comparison, this holds only at a formal level,
in that the limit $F \rightarrow 1$
is not well-behaved. Indeed, the hypotheses of the
theorems in Refs. \cite{Disconzi, DisCzu} require $F>1$ (compare with Eq. 
\ref{F_ent}), as it should be, as
those results ensure a causal dynamics, a feature not shared
by Eckart's theory. In particular, the reader should be aware that,
in light of Eq. (\ref{def_F_general}), setting $F=1$ 
in our equations  corresponds to imposing 
the constraint $p + \rho = \mu$,
a condition that will not hold in general in Eckart's theory. 
We return to this point in Sec. V.  

In any case, despite the necessity of restrictions 
for the applicability of the theorems in Refs. \cite{Disconzi, DisCzu},
 we shall expand our study
 beyond the hypotheses of those theorems.
 We are justified in doing so because such theorems are
 sufficiently general (e.g., they make no symmetry assumption) as 
 to encourage
a detailed study of the physical implications of adopting 
Eq. (\ref{T_vis}). To expand a little more on this point,
we notice that Eq. (\ref{T_vis}) is a particular case of
\begin{align}
\begin{split}
T_{\alpha\beta} &= (p+\rho) u_\al u_\be + p g_{\al\be}
- \zeta \pi_{\al\be} \nabla_\mu C^\mu
\\
& -\vartheta \pi_\al^\mu \pi_\be^\nu (\nabla_\mu C_\nu + 
\nabla_\mu C_\nu)
\end{split},
\label{T_vis_general}
\end{align}
where $\pi_{\al\be} = g_{\al\be} + u_\al u_\be$ and 
$\vartheta$ is the shear-viscosity coefficient.
The equations studied in Refs \cite{Disconzi, DisCzu}
do not have bulk viscosity, but carry the shear term
and make no symmetry assumption on the metric. From the point
of view of the techniques of weakly-hyperbolic systems employed in 
those works, the shear term is the most problematic one
due to the multiple characteristics that arise from  
$\pi_\al^\mu \pi_\be^\nu$. Heuristically, therefore,
when the full stress-energy tensor (\ref{T_vis_general}) is considered,
one expects that causality would fail first due to the presence of the shear term. But, since
causality in the presence of shear viscosity has been
shown, under appropriate assumptions,
in Refs. \cite{Disconzi, DisCzu}, it becomes reasonable 
to suspect that
the simpler case of $T_{\al\be}$ with bulk viscosity only, i.e.,
Eq. (\ref{T_vis}), will also present a good causal behavior.
With these considerations in hand, we point out that here
we are concerned mainly with the applications of 
Eq. (\ref{T_vis}), with the question of the well-posedness
and causality behavior of our equations left for future 
work, where these questions will be addressed in full
detail.

We conclude with some comments on the parametrizations (\ref{T_vis}) and
(\ref{def_F_general}). While at first sight they seem arbitrary,
they are in fact well motivated in light of  known difficulties
when introducing viscosity into General Relativity, difficulties
which are ultimately traced to the lack of a Lagrangian formulation
for viscous fluids. A full discussion is given in Refs. 
\cite{Disconzi, DisCzu} and, to a lesser extent, in Ref.
\cite{Lichnerowicz}. In a nutshell, the adoption of Eq. (\ref{T_vis})
allows one to approach the problem via a traditional point of 
view,
so successful in the study of other  matter models in General Relativity, 
where information about the matter fields is essentially 
contained in the stress-energy tensor, which enters Einstein's 
equations. This is in contrast with the MIS theory, where many other
aspects have to be incorporated, more or less arbitrarily, in the dynamics
(see, e.g., Eq. (\ref{MIS_condition}) below and the discussion that follows).

\section{II. Modified Friedmann equations}

The Friedmann-Robertson-Walker metric with spatially flat
 geometry (in accordance with observations) is
\begin{gather}
ds^2 = -dt^2 + a^2(t)  \Big( dr^2 + r^2 d\theta^2 +
r^2 \sin^2 \theta ~d\phi^2 \Big),
\end{gather}
where $a$ is the scale factor.
In what follows, we will take $a=1$ at the present.

It will be convenient to write $\nabla_\mu C^\mu$ more explicitly.
For the Friedmann-Robertson-Walker metric,
\begin{gather}
\nabla_\mu u^\mu = 3\frac{\dot{a}}{a},
\label{div_u}
\end{gather}
and we find
\begin{gather}
\nabla_\mu C^\mu = \dot{F} + 3 F \frac{\dot{a}}{a}.
\label{div_C}
\end{gather}
The first Friedmann equation then
becomes
\begin{align}
\begin{split}
\dot{H} + H^2 \equiv \frac{\ddot{a}}{a} & 
= - \frac{1}{6} \Big (  3p + \rho 
- 3 \zeta \dot{F} - 9 \zeta F \frac{\dot{a}}{a} \Big )  ,
\end{split}
\label{fri_ddot}
\end{align}
where, as usual, $H \equiv \dot{a}/{a}$ is the  Hubble parameter,
and we have absorbed
a possible cosmological constant into the definition of $\rho$ (with $\rho_\Lambda = - p_\Lambda$).
The second Friedmann equation
remains unchanged, i.e, $H^2 
 = {\rho}/{3}$. 
The evolution of $\rho$ is given by
\begin{align}
\begin{split}
\dot{\rho} + 3(p + \rho) H -
3 \zeta (
\dot{F} +3 F H  )H = 0.
\end{split}
\label{rho_evolution}
\end{align} 
Now we can define an effective pressure via
\begin{gather}
p_{eff} = p + \Pi,
\label{p_eff}
\end{gather}
where $\Pi$ gives the effective change in the pressure due to viscosity.
Then Eqs. (\ref{div_C}) and (\ref{rho_evolution}) give
 \begin{align}
 \begin{split}
 \Pi & =
 -\zeta \nabla_\mu C^\mu  = -\zeta \dot{F} - 3\zeta  F H.
 \end{split}
\label{eff_press}
\end{align}
For the case of Eckart viscosity, one has
$\Pi = - 3\zeta H$, which formally agrees with
Eq. (\ref{eff_press}) upon
setting
$F=1$.

\section{III. Evolution of viscous fluids}

Consider a fluid with an equation of state
\begin{equation}
p = w\rho,
\label{eq_state_w}
\end{equation}
where, e.g., $w=0$ corresponds to nonrelativistic matter, $w=1/3$ to radiation, and $w=-1$ to vacuum energy.
However, we take the most general possible case and allow $w$ to vary with time.

From 
Eq. (\ref{mu_conservation}) and Eq. (\ref{div_u})
one
 immediately 
gets
\begin{gather}
\mu = \mu_0 a^{-3},
\label{mu_sol}
\end{gather}
where $\mu_0$ is the present-day value of $\mu$.
From Eqs. (\ref{def_F_general}), (\ref{p_eff}), (\ref{eff_press}), (\ref{eq_state_w}),  and
 (\ref{mu_sol}) we find
 \begin{gather}
 p_{eff} = p 
 - \frac{ \zeta \dot{w} \rho  }{\mu_0 a^{-3}}
 - \frac{ \zeta(1+w) \dot{\rho} }{\mu_0 a^{-3}}
 - \frac{ 6 \zeta(1+w) \rho H}{\mu_0 a^{-3}}.
\label{eff_press_1}
\end{gather}
Since
$\dot \rho + 3H(\rho + p_{eff}) = 0$, we can eliminate 
$\dot \rho$ from Eq. (\ref{eff_press_1}):
\begin{align}
\begin{split}
p_{eff} &= \left(1 - \frac{3H \zeta (1+w)}{\mu_0 a^{-3}}\right)^{-1}
\\
& \times
\left(p - \frac{\zeta \dot w \rho}{\mu_0 a^{-3}} - \frac{3 H \zeta (1+w)\rho}{\mu_0 a^{-3}}\right).
\end{split}
\nonumber
\end{align}
Then the effective equation of state parameter, $w_{eff} \equiv p_{eff}/\rho$, which gives $\rho(a)$ via
\begin{equation}
\label{drhoda}
\frac{d\ln\rho}{d\ln a} = -3(1+w_{eff}),
\end{equation}
is given by
\begin{equation}
\label{w_eff}
w_{eff} = \frac{w \mu_0 a^{-3} - {\zeta \dot w} - {3 H \zeta (1+w)}}
{\mu_0 a^{-3} - {3H \zeta (1+w)}}.
\end{equation}
Lacking detailed knowledge of the functional form of $\zeta$, we will follow earlier treatments
and simply take $\zeta$ to scale as an undetermined power of the density, namely,
$\zeta = \zeta_0 \rho^\alpha$. 
Also, for simplicity, we will assume from now on that $w$ is constant.

There are several special cases of interest.  
When $w=-1$ (vacuum energy), viscosity has no effect.  This
is clear from the definition of $F$ (Eq. \ref{def_F_general}), which shows that $F=0$ for $p + \rho
=0$, so there is no viscosity in our model.
The opposite extreme, a stiff equation of
state with $w=1$, also yields no effect on the density evolution, as Eq. (\ref{w_eff}) gives
$w_{eff} = w = 1$ in this case. This can be understood as follows. For an ideal fluid,
it is possible to show that stiffness is equivalent 
to $\nabla_\al C^\al = 0$ \cite{LichnerowiczBook}. In fact, this
is one of the motivations to introduce $F$: in Newtonian fluids,
the incompressibility condition is assured by the vanishing 
of the divergence of the velocity. A stiff fluid is the relativistic analogue 
of incompressible Newtonian fluids, and, therefore, we would like a similar
divergence-free condition to hold, except that now we speak of a
four-divergence. This is possible if we consider the 
dynamic velocity $C$ instead of the ordinary four-velocity $u$. For perfect fluids,
considering the stress-energy tensor uniquely in terms of $u$ or $C$ leads
to the same results. However, as we have stressed,
there is a fundamental difference in which of these quantities we take 
as defining the viscous part of $T_{\al\be}$.
On the other hand, the feature of a fluid being stiff should not depend
on whether viscosity is present, exactly in the same way that incompressibility
for Newtonian fluids is defined by a divergence-free condition in both
the Euler and the Navier-Stokes equations. But under the assumption of 
a stiff fluid, i.e., $\nabla_\al C^\al = 0$, the bulk term drops out of
Eq. (\ref{T_vis}), consistent with the previous behavior when $w=1$.
It is essential
to stress, however, that this absence of viscosity effects is a consequence
of the symmetry of the problem, and not of the general model \cite{Disconzi}
on which we base our equations. Indeed, in a Robertson-Walker space-time,
the only allowed contribution to viscosity comes from the bulk term, but
in a general space-time, shear viscosity will be present when the
fluid is stiff.

Next, consider the case where the viscous fluid with constant $w$ dominates the expansion, so that
$H^2 = \rho/3$.  Then Eqs. (\ref{drhoda}) and (\ref{w_eff}) give:
\begin{equation}
\label{rhoevol}
\frac{d\ln\rho}{d\ln a} = -3(1+w)
\left(\frac{1 - 2 \sqrt{3}(\zeta_0/\mu_0) a^3 \rho^{\alpha + 1/2}}
{1 - \sqrt{3}(1+w)(\zeta_0/\mu_0) a^3 \rho^{\alpha + 1/2}}\right) .
\end{equation}
We can make some general qualitative arguments regarding the density
evolution in this case.  First assume that the viscosity
is negligible at early times, so that the second terms
in both the numerator and denominator of Eq. (\ref{rhoevol}) are $\ll 1$
when $a \ll 1$.  For negligible viscosity, $\rho$ evolves
as $a^{-3(1+w)}$, so both viscosity correction terms scale as
$a^{3 - 3(1+w)(\alpha + 1/2)}$.  Thus, the viscosity correction to the equation of state
will grow with time as long as
$
\alpha < \frac{1-w}{2(1+w)}.
$
The result will be (as expected) a value of $\rho(a)$ that decreases
more slowly than in the standard non-viscous case.  Then, when $a$
reaches the value for which $\mu_0 a^{-3} < 2 \sqrt{3}\zeta_0 \rho^{\alpha + 1/2}$,
the value of $w_{eff}$ drops below $-1$, and $\rho$ begins to increase with $a$.
This phantom evolution inevitably results in a future singularity \cite{Caldwell}.
We can see from Eq. (\ref{rhoevol}) that this singularity is reached when
$\mu_0 a^{-3} = \sqrt{3}(1+w)\zeta_0 \rho^{\alpha + 1/2}$, at which point $w_{eff} \rightarrow -\infty$.
(For a discussion of future singularities with a different, more general set of viscosity-motivated
modifications to the effective pressure, see Refs. \cite{Od1,Od2,Od3}).
Note that a value of $w_{eff} < -1$, while puzzling from a theoretical perspective, is not ruled
out by current cosmological data and may even be observationally
favored over $w_{eff} > -1$ \cite{Hinshaw,Ade,Betoule}.  Our model also provides an elegant
way for the dark matter to ``cross the phantom divide," evolving from $w_{eff} > -1$
to $w_{eff} < -1$, something which is notoriously difficult to achieve in more conventional 
dark energy models.

Explicit solutions to Eq. (\ref{rhoevol}) can be obtained for several special cases.
Consider first $\alpha = -1/2$.  This value has no special significance, but the
analytic solution illustrates some of these qualitative arguments.  For this special case, the solution to
Eq. (\ref{rhoevol}) is
\begin{equation}
\rho \propto a^{-3(1+w)}(1-B a^3)^{w-1},
\end{equation}
where
$B = \sqrt{3}(1+w)(\zeta_0/\mu_0)$.
This solution exemplifies our earlier qualitative arguments.  For $Ba^3 \ll 1$,
we simply have standard non-viscous evolution.  But when $Ba^3$ increases to $O(1)$, $\rho$ decreases
more slowly than in the non-viscous case.  At $Ba^3 = (1+w)/2$, the value of $w_{eff}$ decreases below
$-1$, and the density begins to increase with the scale factor (phantom-like behavior),
ultimately approaching $-\infty$ as $Ba^3 \rightarrow 1$.

A more complex implicit solution is obtained for the value $\alpha = -7/2$, namely
\begin{gather}
\rho^\frac{1}{7}
a^{\frac{6}{7} - \frac{3(1-w)}{7(4+3w)}}
\Big[ (4+3w) \rho^3 - \frac{7 \sqrt{3} \zeta_0}{\mu_0}(1+w) a^3 
\Big]^\frac{1-w}{7(4+3w)}
\nonumber
\\
= \text{constant}.
\nonumber
\end{gather}
Another exact solution is found for $\alpha = -5/6$ and
$w = -2/3$, for which
\begin{equation}
a \rho - \frac{ \sqrt{3} \zeta_0 }{2 \mu_0} a^4 \rho^\frac{2}{3} = \text{constant}.
\end{equation}

These solutions can be compared to the evolution in
 the case of conventional Eckhart theory.
Repeating the above arguments with Eq. (\ref{T_N}), so that $p_{eff} = p - 3\zeta H$, leads to a solution for $\rho(a)$ for arbitrary
$\alpha$ and constant $w \ne -1$, namely
\begin{equation}
\label{Eckrho}
\rho = \left[\sqrt{3} \frac{\zeta_0}{1+w} + C a^{3(1+w)(\alpha - 1/2)}\right]^{2/(1-2\alpha)},
\end{equation}
for $\alpha \ne 1/2$.  (The case $\alpha = 1/2$ simply yields a constant difference between $w_{eff}$
and $w$).
In Eq. (\ref{Eckrho}), $C$ is a constant that can be set to give $\rho = \rho_0$ at $a=1$.
For $\alpha < 1/2$, viscosity is subdominant at early times, so
$\rho \propto a^{-3(1+w)}$, while $\rho$ approaches 
a constant at late times.
In contrast, as seen above, our model gives a density  
that evolves to phantom-like behavior at late times.

\section{IV. Entropy production}
Using Eqs. (\ref{T_vis}), (\ref{def_F_general}), (\ref{mu_conservation}),
and the conservation law $u^\be \nabla^\al T_{\al\be} = 0$,  we find
\begin{align}
\begin{split}
u^\al \partial_\al p  & -\mu u^\al\partial_\al F  
\\
=
-\zeta F (\nabla_\al u^\al)^2 & - \zeta \nabla_\al u^\al u^\beta \partial_\beta F .
\end{split}
\label{cons_F_p_u}
\end{align}
Since $\rho = \mu(1 + e)$,
where $e$ is the specific internal energy, we see that $p + \rho = 
\mu(1 + e + \frac{p}{\mu})$, and thus comparison with Eq. (\ref{def_F_general}) yields
 \begin{gather}
F = 1 + e + \frac{p}{\mu} .
\label{F_ent}
\end{gather}
In particular, we see that $F$ is the relativistic specific enthalpy
of the fluid, and furthermore, we can identify the dynamic velocity $C^{\mu}$ of Eq. (\ref{T_vis})
as the enthalpy current. (For a discussion, see e.g.,  Ref. \cite{RZ}.)
From Eq. (\ref{F_ent}),
$dF = de + p d\left(\frac{1}{\mu}\right) + \frac{1}{\mu} dp$,
which combined with the first law, i.e., $
T ds = de + p d\left(\frac{1}{\mu}\right)$,
produces
$-\mu T ds = - \mu dF + dp$, or yet
$-\mu T \partial_\al s = - \mu \partial_\al F + \partial_\al p$.
Here, $T$ is the temperature and $s$ the specific entropy. 
Contracting the last equality with  $u^\al$, combining with
Eq. (\ref{cons_F_p_u}), and invoking Eq. (\ref{div_u}), finally produces
\begin{gather}
\mu T \dot{s} = 3 \zeta H( \dot{F} + 3FH).
\label{entropy_change}
\end{gather}
For the sake of brevity, we shall restrict ourselves to the case
$p = w \rho$, with $w$ constant, as in the previous section, although 
our conclusions hold under other conditions.
Let us also suppose in this section
 a general behavior of the form $\rho \propto a^\beta$ at lowest order,
which is consistent with the discussion of Sec. III and much 
of  the intuition drawn from standard cosmology.
As $\mu \geq 0$, $T \geq 0$, and in the cases of 
interest we can assume $H > 0$ and  $\zeta\geq 0$, we immediately see that, 
if $\beta \geq -6$, which covers a wide range of possible models,
we obtain
$\dot{s} \geq 0$, in accordance with the second law of thermodynamics. 
Notice that
 equality happens when $\zeta = 0$ (i.e., no viscosity).

We can also analyze the entropy current 
$S_\al \equiv s\mu u_\alpha$.
Using Eqs. (\ref{mu_conservation}) and (\ref{entropy_change}), we find at once
that
\begin{gather}
T \nabla_\al S^\al = 3\zeta H( \dot{F} + 3FH),
\label{s_current}
\end{gather}
and, once again, that the second law,
$\nabla_\al S^\al \geq 0$, is satisfied under the same conditions
as above.

These results should be contrasted with models based on 
the MIS theory \cite{Maartens}, where $\nabla_\al S^\al \geq 0$ does not follow
from the MIS stress-energy tensor and simple
scaling arguments for the thermodynamic quantities,
but rather is dynamically imposed along with  a redefinition of 
$S_\al$.

More precisely, let $\widetilde{S}^\al$ be the entropy current 
as in the MIS theory, i.e, $\widetilde{S}_\al = S^\al - \frac{\tau \widetilde{\Pi}^2}{2T \xi} u_\al$, where  $\xi \geq 0$ is the bulk viscosity coefficient,
$\tau \geq 0$ is the relaxation coefficient for transient bulk viscous effects,
and $\widetilde{\Pi}$ 
is the MIS bulk viscous stress. From Eq. (\ref{s_current}), it follows that
\begin{align}
\begin{split}
\nabla_\al \widetilde{S}^\al 
& =
\frac{3H}{T}\Big ( \frac{\zeta}{T} \dot{F} + 3\frac{\zeta}{T} F H
- \frac{\tau \widetilde{\Pi}^2}{2T \xi} \Big)
\\
&
- \frac{\tau \widetilde{\Pi}^2}{2T \xi}\Big( 
2 \frac{\dot{\widetilde{\Pi}}}{\widetilde{\Pi}} + \frac{\dot{\tau}}{\tau}
- \frac{\dot{T}}{T} - \frac{\dot{\xi}}{\xi} \Big ).
\end{split}
\label{div_S_MIS}
\end{align}
In the MIS formulation, it is imposed that 
\begin{gather}
\widetilde{\Pi} + \tau \dot{\widetilde{\Pi}} = -3\xi H - \frac{1}{2} \tau \widetilde{\Pi}
 \Big( 3H
+ \frac{\dot{\tau}}{\tau}
-\frac{\dot{\xi}}{\xi} 
-\frac{\dot{T}}{T}
 \Big ).
\label{MIS_condition}
\end{gather}
Combining  Eq. (\ref{MIS_condition}) with Eq. (\ref{div_S_MIS}) gives
\begin{align}
\begin{split}
\nabla_\al \widetilde{S}^\al  = 
\frac{3H \zeta}{T}  (  \dot{F} + 3 F H )
+
 \frac{3 H \widetilde{\Pi} }{T }  + \frac{\widetilde{\Pi}^2}{T \xi}.
\end{split}
\label{div_S_MIS_2}
\end{align}
Typically, $\widetilde{\Pi}$ is negative, and thus 
the sum of the last two terms in Eq.
(\ref{div_S_MIS_2}) will be non-negative, implying
$\nabla_\al \widetilde{S}^\al \geq 0$, if 
$\widetilde{\Pi} \leq -3H \xi$, which will be satisfied
for $\xi$ sufficiently small. This can be relaxed since, under
our assumptions, the first term in $\zeta$ gives a positive contribution.

Although this analysis remains very qualitative, the 
point here is that if we insist
on defining the entropy current by $\widetilde{S}^\al$ and 
employ the assumptions of the MIS theory, 
while at the same time adopting Eq. (\ref{T_vis}),
we still
obtain that $\nabla_\al \widetilde{S}^\al \geq 0$
under reasonable conditions.
In any case, it is important to remember that Eq. (\ref{MIS_condition}) 
is, in a sense,
arbitrary. It is adopted in that it constitutes the simplest condition, linear
in $\widetilde{\Pi}$, that enforces 
$\nabla_\al \widetilde{S}^\al \geq 0$, but other conditions
can be imposed. For instance, in our context,
$\nabla_\al \widetilde{S}^\al \geq 0$ will hold if 
$\widetilde{\Pi}$ satisfies, instead of Eq. (\ref{div_S_MIS}),
the  equation
$
\dot{\Pi} + \frac{1}{2}  ( 
\frac{\dot{\tau}}{\tau}
- \frac{\dot{\xi}}{\xi}
-
\frac{\dot{T}}{T}
+ 4H   ) \Pi = 0
$. Yet other evolution equations for $\widetilde{\Pi}$ can be devised. 
For instance, one may set the entire right-hand side of Eq. (\ref{div_S_MIS_2})
equal to a positive combination of the thermodynamic quantities.

The aim of these considerations is to emphasize 
the flexibility of our approach.
While we showed above that $\nabla_\al S^\al \geq 0$ and $\dot{s} \geq 0$
follow naturally from the equations of motion under simple assumptions, 
one can still adopt a MIS-like
 point of view
when desirable.
In this sense, models based on Eq. (\ref{T_vis}) can be viewed as mixed between
the traditional approach (on which Eckart's theory is based), and the 
Extended Irreversible Thermodynamics (on which MIS's theory is based).

\section{V. The limit $F\rightarrow 1$}

Here we explore in more detail the behavior, mentioned in the introduction,
 of solutions when
$F \rightarrow 1$. We start by
pointing out that if we restore the units in (\ref{F_ent}), we have
$F = 1 + c^{-2} (e + p/\mu)$, where $c$ is the speed of light.
Thus, $F = 1 + O\left(\frac{1}{c^2}\right)$, which justifies the notion
that $F$ gives a relativistic correction to $u$, as initially remarked.

To explore the limit $F\rightarrow 1$, it is instructive to suppose 
that we are working in the traditional thermodynamic setting, where
the pressure $p$ and the specific energy are non-negative. In this case,
setting $F=1$ in (\ref{F_ent}) implies $p=0=e$. This corresponds
to the special case of pressureless dust, for which $w=0$.
The first law of thermodynamics
now gives $T ds = 0$, which combined with (\ref{entropy_change}) 
leads to $\zeta  = 0$, provided that $T > 0$ and $H > 0$. 
We can understand this in two
ways. Generally, $\zeta$ is a function of the thermodynamic quantities, and 
hence a function of $F$, $\zeta = \zeta(F)$. Therefore, we see that $\zeta(F)
\rightarrow  0$ as $F \rightarrow 1$. 
This is consistent with the idea that in the limit of zero pressure
the interaction rate due to particle collisions
should go to zero (for finite interaction lengths),
so in that sense there should be no dissipation.
We can, however, consider a second possibility, namely, the case where $\zeta$ is constant and non-zero. In this 
situation, passing to the limit $F \rightarrow 1$ gives a contradiction. Keeping
in mind that (\ref{entropy_change}) relies on the equations of motion,
this means that, although
we can obtain well-behaved solutions without superluminal signals for $F>1$,
these results do not pass to the limit; in other words, the limit of solutions
is not, in general, a solution, when $\zeta$ is constant.

\section{VI. Final comments}

The model for relativistic viscosity introduced in Ref. \cite{Disconzi} combines the
advantages of Eckart's model with those of the MIS theory.
It is nearly as simple as Eckart viscosity but does not have the causality problems of that model.
It is much simpler than the MIS theory, and, like the MIS theory, it is plausible to conjecture
that the model is causal for all physical systems of interest, although at present
neither the model discussed here nor the MIS model can be rigorously {\it proven} to be causal under all
possible circumstances.  Acceptable  thermodynamic behavior (i.e.,
agreement with the second law of thermodynamics) under reasonable circumstances
is also achieved. 

This model also yields a number of interesting
results when applied to cosmological fluids.  One appealing property is that it automatically
reduces to zero viscosity for both vacuum energy and a stiff fluid.  We find that constant-$w$ fluids
can produce a future singularity for a wide range of parameter choices. 

All of these features make Eq. (\ref{T_vis}) a promising candidate 
for a viscous stress-energy tensor in cosmology, inviting further investigation of 
the model.

\section{Acknowledgments}

We thank J. Barrow, S. Odintsov, and I. Brevik for helpful comments on the manuscript.
T.W.K. and R.J.S. are supported in part by the DOE (DE-SC0011981).
M.M.D. is supported in part by the NSF (grant 1305705).

\newcommand\AJ[3]{~Astron. J.{\bf ~#1}, #2~(#3)}
\newcommand\APJ[3]{~Astrophys. J.{\bf ~#1}, #2~ (#3)}
\newcommand\apjl[3]{~Astrophys. J. Lett. {\bf ~#1}, L#2~(#3)}
\newcommand\ass[3]{~Astrophys. Space Sci.{\bf ~#1}, #2~(#3)}
\newcommand\cqg[3]{~Class. Quant. Grav.{\bf ~#1}, #2~(#3)}
\newcommand\mnras[3]{~Mon. Not. R. Astron. Soc.{\bf ~#1}, #2~(#3)}
\newcommand\mpla[3]{~Mod. Phys. Lett. A{\bf ~#1}, #2~(#3)}
\newcommand\npb[3]{~Nucl. Phys. B{\bf ~#1}, #2~(#3)}
\newcommand\plb[3]{~Phys. Lett. B{\bf ~#1}, #2~(#3)}
\newcommand\pr[3]{~Phys. Rev.{\bf ~#1}, #2~(#3)}
\newcommand\PRL[3]{~Phys. Rev. Lett.{\bf ~#1}, #2~(#3)}
\newcommand\PRD[3]{~Phys. Rev. D{\bf ~#1}, #2~(#3)}
\newcommand\prog[3]{~Prog. Theor. Phys.{\bf ~#1}, #2~(#3)}
\newcommand\RMP[3]{~Rev. Mod. Phys.{\bf ~#1}, #2~(#3)}

\end{document}